# Review—The Development of Wearable Polymer-Based Sensors: Perspectives

View the article online for updates and enhancements.

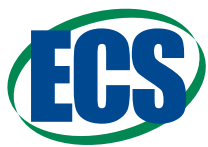

# Review—The Development of Wearable Polymer-Based Sensors: Perspectives

Christian Harito,[1,2] Listya Utari,[3,4] Budi Riza Putra,[5,6] Brian Yuliarto,[3,4,z] Setyo Purwanto,[7] Syed Z. J. Zaidi,[8] Dmitry V. Bavykin,[8] Frank Marken,[5,*] and Frank C. Walsh[8]

[1]*Department for Management of Science and Technology Development, Ton Duc Thang University, Ho Chi Minh City, Vietnam*
[2]*Faculty of Applied Sciences, Ton Duc Thang University, Ho Chi Minh City, Vietnam*
[3]*Advanced Functional Materials (AFM) Laboratory, Engineering Physics, Institut Teknologi Bandung, 40132 Bandung, Indonesia*
[4]*Research Center for Nanosciences and Nanotechnology(RCNN), Institut Teknologi Bandung, 40132, Bandung, Indonesia*
[5]*Department of Chemistry, University of Bath, Claverton Down, BA2 7AY Bath, United Kingdom*
[6]*Department of Chemistry, Faculty of Mathematics and Natural Sciences, Bogor Agricultural University, Bogor, West Java, Indonesia*
[7]*Center for Science and Technology of Advanced Materials–National Nuclear Energy Agency Kawasan PUSPIPTEK Serpong, Tangerang Selatan 15314, Indonesia*
[8]*Energy Technology Research Group, Faculty of Engineering and Physical Sciences, University of Southampton, SO17 1BJ, Southampton, United Kingdom*

The development of smart polymer materials is reviewed and illustrated. Important examples of these polymers include conducting polymers, ionic gels, stimulus-response be used polymers, liquid crystalline polymers and piezoelectric materials, which have desirable properties for use in wearable sensors. This review outlines the mode of action in these types of smart polymers systems for utilisation as wearable sensors. Categories of wearable sensors are considered as tattoo-like designs, patch-like, textile-based, and contact lens-based sensors. The advantages and disadvantages of each sensor types are considered together with information on the typical performance. The research gap linking smart polymer materials to wearable sensors with integrated power systems is highlighted. Smart polymer systems may be used as part of a holistic approach to improve wearable devices and accelerate the integration of wearable sensors and power systems, particularly in health care.

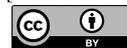

There is considerable interest in the development of wearable technology to enhance the functionality of devices such as smart watches, medical devices, smart glasses, wrist bands, eye wear and smart clothing. Functions including sensing, energy harvesting, luminescence and generation of thermos-electricity have been embedded in smart devices which can be applied for healthcare,[1] safety,[2] fashion, fitness,[3] data gathering, leisure and entertainment. Growing interest in healthcare has encouraged research in the development and refining of such smart devices, particularly for biosensing,[4] beyond the current technology used in smart watches. Wearable sensors with real-time monitoring and diagnostic system could become a key factor in transformation the healthcare industry.

Wearable sensors have been used for monitoring elderly and senior[5] patients, those with chronic diseases[6] and athletes. The devices may be deployed in daily life to assist doctors and nurses to monitor patients at home or prevent diseases as well as to study the health status of babies. Comfortable wearable sensors are poised to see increasing use in lifestyle products. Polymers have a long history as wearable materials, which began with clothing and textiles. In ancient Egypt, natural polymers, such as flax fibre spun from the stem of a plant, was used for linen clothing from the end of Neolithic period, before 3100 BC.[7] During the Han dynasty, around 114 BC, silk clothing became popular and spread around the continent by the silk road trade route, aiding the development of great civilizations in China.[8] Smart textiles based on polymers show promise incorporating new additional sensor functionalities in next generation clothing products. The versatility of polymers allows them to be synthesised in several morphologies, including gels,[9] liquid crystal polymers[10] and elastomers, extending the morphology of wearable sensors into patch-like sensors even micrometre thick, tattoo-like sensors.[11] The wearable sensor itself has been developed form its first invention around 1980 in the form of chest strap wireless electrocardiography.[12] The chest strap sensor shows the importance of flexibility in wireless sensor hence the utilisation of flexible material such as polymer is essential especially as substrate. The first wearable sensor has also inspired researchers to utilise textile for enhancing the flexibility of electronics. The development of textile-based sensor is even extended to wearable computing devices,[13] transistors,[14] and energy storages such as batteries[15,16] and fuel cells.[17] Besides the textile-based sensor, the patch-like sensor was also developed in 1992 as transdermal alcohol vapour sensor.[18] However, there was 0.5–2 h delay in detection using body vapour, which later motivates researchers to use another source of detection such as sweat, tears, saliva, and interstitial fluid. In the last decades, the development of wearable sensors has increased exponentially especially in tattoos and contact lens sensors in which the role of polymers in each sensor is discussed in this review.[19–22] The key milestones of the wearable sensor including significant papers and reviews are highlighted in Fig. 1.

Recently, several reviews on wearable sensor have been published which emphasise a wearable alcohol sensor,[28] a gloved-based chemical sensor focussed on transforming benchtop chemical analysis into fingertip analyser,[29] wearable biofuel cells[30] and wearable chemical sensors. A comprehensive review on chemical and electrochemical sensors[31] has been published. As the design of wearable sensors (e.g., tattoo, textile and contact lens types) has exponentially increased over the last two decades, it is important to acknowledge the significance of polymer materials in each device, since the introduction of wearable sensors around 1980, considering the fundamental role of the polymer as a flexible substrate. In this review, the discussion is arranged according to the polymer material and design perspectives. Types of polymer used in wearable sensor are discussed. The term "wearable sensor" in this review is used to mean a sensor that measures body signal parameters without external interactions. The relationship of polymers properties with the sensing system are outlined. Synthesis methods of several wearable sensor morphology based on polymer are mainly centred on textile-like, patch-like, tattoo-like and contact lens sensors which are the most common types in wearable sensors. The application of wearable sensors is based on the several measurements such as chemical, mechanical, electrical, optical, and its combinations. The integration of a wearable sensor into a power system, e.g., via a biofuel cell, may become an important research direction in continuous body monitoring, so the future design of wearable sensors is considered.

*Electrochemical Society Member.
[z]E-mail: brian@tf.itb.ac.id

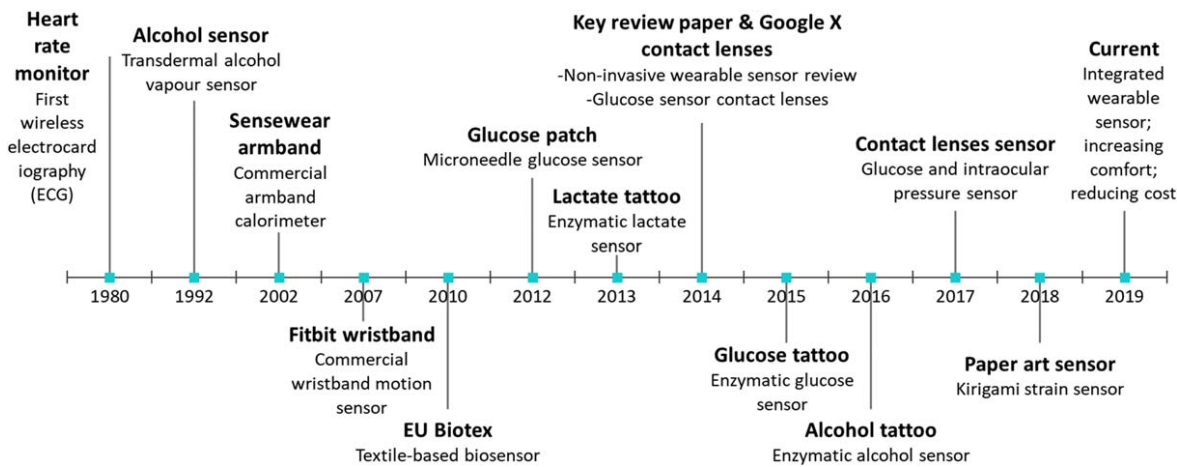

**Figure 1.** The key milestones of the wearable sensor. Heart rate monitor,[12] Alcohol sensor,[18] EU Biotex,[23] Glucose patch,[24] Lactate tattoo,[25] Key review paper,[26] Glucose tattoo,[19] Alcohol tattoo,[20] Contact lens sensor,[21,22] paper art sensor.[27]

### Types of Polymer

Although the first synthetic polymer, Bakelite was synthesised in 1907, the molecular nature of polymers was poorlyunderstood at that time. In 1922, Staudinger laid down the foundation of macromolecular chemistry which led to a Nobel Prize in 1953. It became well appreciated that polymers had high relative molecular mass and contained repeating unit of molecules, creating a long chain of atoms connected by covalent bonding. The continued development of functional and intelligent polymers has led to wearable sensors. Polymers can be used as substrate, electrodes, and active materials in such devices. The flexibility and toughness of polymers is important in their use as substrates. Electrically conducting polymers or composites and stimulus-response polymers can be used as active electrodes. The types of polymer used in wearable sensors are outlined in Table I.

Liquid crystalline polymers (LCPs) retain an ordered structure in liquid or molten form at certain temperature, pressure, and concentration.[40] There are two types of LCP. Lyotropic types arise from solvation of polymer, while thermotropic LCPs are created by heating. Thermotropic liquid crystals were first observed by Friedrich Reinitzer in 1888, when two different colours and melting temperature are shown during transition from liquid to solid of cholesteryl benzoate.[41] The finding was supported by Otto Lehmann, who characterised the liquid crystal and found multiple small crystalline formation in the molten cholesteryl benzoate.[42] The in-between phase of liquid and solid was latter termed the mesophase, in which the ordered individual molecules within liquid or molten form termed *mesogens*. The mesophase structures are generally categorised into nematic and smectic depending on its orientation and the degree of order.[43] The order of molecules is increased from isotropic liquid, nematic, smectic phase, to crystalline solid in Fig. 2. Each phase has particular physical characteristics (e.g., optical, electromagnetic), which inspires researcher to control the switching phases with electricity and utilises it for liquid crystal display (LCD). The same principles can be used for wearable sensor as well, which it can be controlled by body signal parameters such as electrical, chemical, thermal, and pressure. The liquid crystal polymer may retain its orientation during cooling and creates strong and aligned polymer as a substrate in wearable sensors.

The supramolecular structure of liquid crystal polymer can be divided into main-chain, side-chain polymer, and polymer network in Fig. 3.[44] The changes in the degree of order and phase transitions are developed based on the controlled formation of non-covalent bonding arising from hydrogen, halogen, ionic bond, and charge-transfer interactions. In the main-chain structure, the ratio of co-polymers is the key to define the mesogenic phase and phase transition behaviour through self-assembly. The degree of cross-linking in liquid crystal side-chain polymer or polymer network allows the phase transition to occur in the form of gel or elastomer, which can be observed by selective shrinkage. The stimulus-response behaviour can be induced in polymer structure by attaching functional moieties to the *mesogen*. For example, photo-chromic moieties such as spiropyran, azobenzene, diarylethene, and spiroxazine exhibit photo-sensitivity[45] while π conjugated moieties provide electroactive properties[46] and ionic or moieties with salt complex incorporate ion sensitive properties.[47] Owing to the high degree of hydrogen bonding, the LCP is very sensitive to water and can be used as an active material in humidity sensors. When the LCP is cooled down into a solid, anisotropic polymer crystals are formed showing a response to the force in a particular direction.

Polymer gels are another stimulus-response polymer often used as an active material in wearable sensors.[48] Being able to absorb and desorb fluid, the physical properties of gel can be easily controlled by displacing the fluid. Temperature-responsive gels are characterised by a lower critical solution temperature (LCST) and upper critical solution temperature (UCST) in Fig. 4, which shows the transition between gel and sol phases due to its solubility.[49] At the LCST, the insoluble gel starts to form above the critical temperature and soluble sol occurs below the critical temperature. UCST, the sol is created above the critical temperature and gel forms below this temperature.

Polymer gels shrink within the temperature range of the soluble sol and swell when they reach the temperature of the insoluble gel. Poly(N-isopropylacrylamide) (PNIPAM) has LCST around 33 °C in water, which is the common polymer gel for sensing human body.[50] The LCST from 25–100 °C can be tuned by copolymerisation introducing hydrophobic properties into the gel and by varying molecular weight.[51,52] In addition to temperature response, some polymer gels may also respond to a chemical stimulus including pH, enzymes, ions, and chemical vapours in which the sensitivity can be controlled through crosslinking density, functional groups and hydrogen bonding of the hydrogel.[53] The pH-dependent swelling or shrinking behaviour of hydrogel is attributed to ionisable pendant groups in polymer backbone while its responsiveness defines by the combination of degree of crosslinking and polymerisation, hydrophobicity/hydrophilicity, as well as the concentration, charge, and *pKa* of the ionizable groups.[54] For biomedical application, the swelling and shrinking behaviour are useful for controlled expulsion of drug, regulating enzyme function and gene expression by binding and releasing target protein, as well as adjusting the tissue adhesion properties by changing the hydrophobicity/hydrophilicity.[55] Functionalised hydrogel can also be used for fluorescent imaging, X-ray microtomography, magnetic resonance imaging when incorporated with certain contrast agent such as organic dyes, inorganic fluorophores, fluorescent proteins for fluorescent imaging, barium or iodine salts for X-ray microtomography and gadolinium or super paramagnetic iron-oxide nanoparticles (SPION) for magnetic resonance imaging.[56] Ionic polymer gels are electrically responsive, containing the electrolyte in a polymer matrix.[57] By combining ionic gel and pressure sensitive polymer such as piezoelectric polymer and dielectric elastomer, a highly sensitive pressure or strain sensor with up to 3.1 nF kPa$^{-1}$ and 500% extension in stretching can be made,[58] which are desirable properties for electrodes in a strain sensor. The piezoelectric or dielectric elastomer can be used as an active material in which its capacitance can be changed during bending or stretching.

**Table I.** Types of polymer and their application in wearable sensors.

| Type of Polymer | Description | Application |
| --- | --- | --- |
| Liquid crystalline polymers (LCP) | A polymer which under suitable conditions of temperature, pressure, and concentration, are thermodynamically stable to form liquid crystal mesophase, which combines the properties of liquid (e.g., ability to flow) and crystalline solid (e.g., anisotropic physical properties).[32] | As substrate, when cooled to an anisotropic solid. An active material can be induced by chemical, thermal and pressure changes.[33] |
| Polymer gels | A non-fluid polymer network that can be expanded throughout its volume by fluid.[34] | Active material induces by chemical, thermal, electrical from body or to be used as electrode (e.g., ionic gels). |
| Intrinsically conducting polymers | An electrically conducting macromolecules consist of conjugated sequences of double bonds or aromatic group, which undergo redox transformation creating charge-transfer complexes by doping.[35] | Active material: to be used as electrode in electrochemical sensor or as an organic semiconductor in transistor-based sensors. |
| Polymer composites | A combination of polymer and filler which usually provide a better performance than pure polymer.[36] | Active material used as electrode or stimulus-response polymer in resistance-based sensors. |
| Piezoelectric polymer | A polymer that converts mechanical energy to electricity and *vice versa*. | Active materials as pressure sensor such as polyvinylidene fluoride (PVDF) and polylactic acid (PLA). |
| Elastomers | A polymer that possess rubber-like elasticity.[37] | Active material (e.g, dielectric elastomers). Substrate (e.g., silicone rubber) |
| Thermoplastic polymers | A polymer that can undergo thermally reversible phase transition between solid and liquid.[38] | Substrate |
| Thermosetting polymers | An insoluble polymer network that made by irreversible curing of viscous prepolymer.[38] | Substrate[39] |

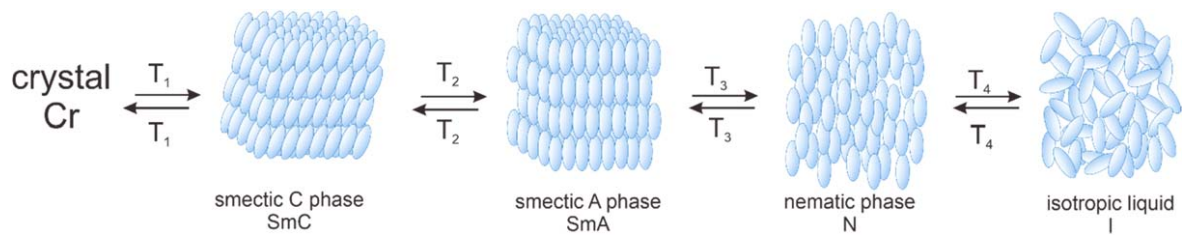

**Figure 2.** Phase transformation of a liquid crystal from crystalline to isotropic liquid.

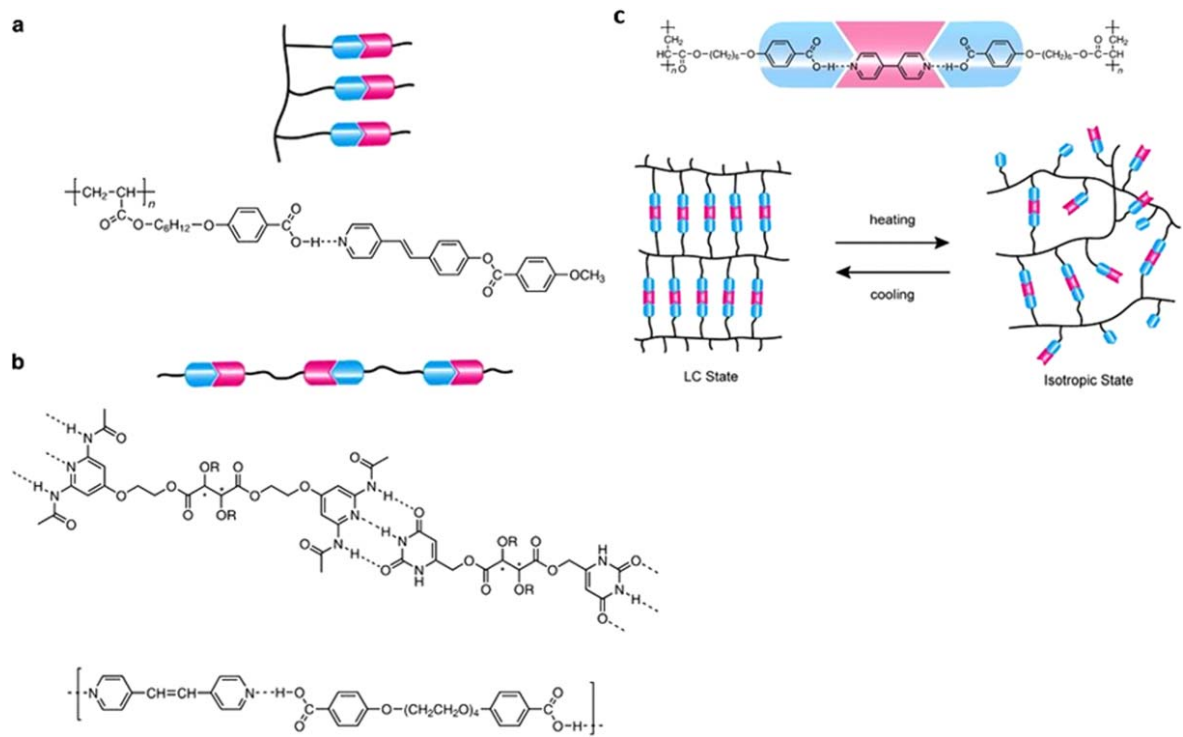

**Figure 3.** An example of a liquid crystal polymer via a non-covalent bonding phase transition. (a). side-chain, (b). main chain, (c). polymer network.[44]

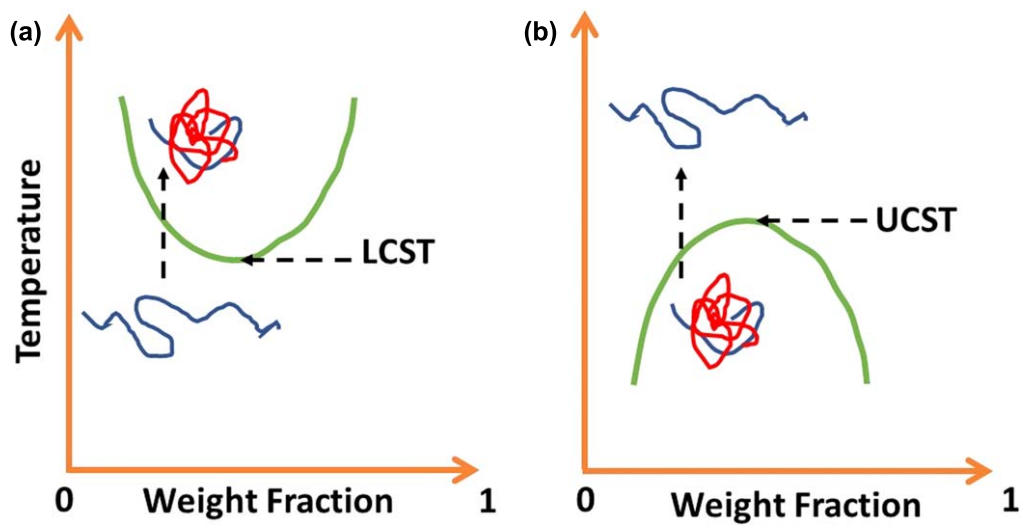

**Figure 4.** The solubility diagram of temperature-responsive polymer gel with critical gel to solution temperature and *vice versa*. (a). upper critical solution temperature (UCST). (b). lower critical solution temperature (LCST).[49]

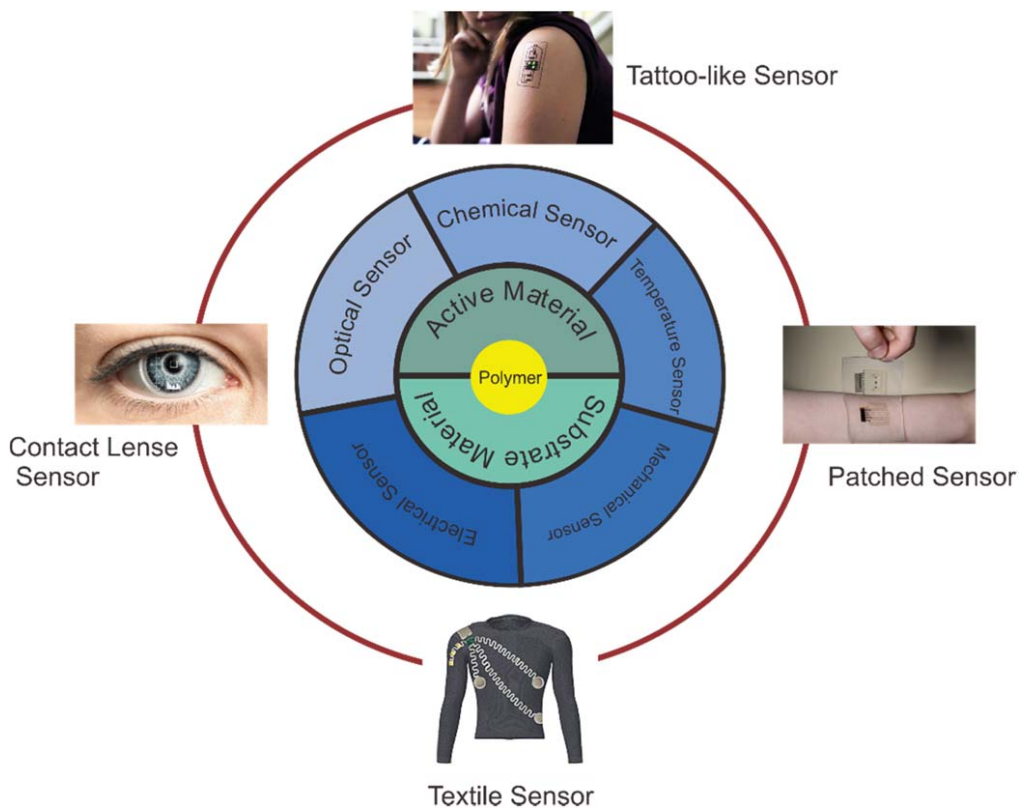

**Figure 5.** The polymer as a substrate or active material (electrode and stimulus-response) for chemical and physical sensing in wearable devices.

Many challenges to body comfort noise, and durability can be posed by sensors. Conventional wet electrods based on Ag/AgCl and gels may cause skin irritation and decreasing performance when it dries.[59] Although the dry electrode based on metal is durable, it may be inconvenient, painful, or cause discomfort. Conducting polymers are becoming more popular as dry electrodes.[60] Conducting polymers can be achieved by using intrinsically conducting polymer or by mixing polymer with conductive filler such as graphene/chitosan composite.[61] Intrinsically conducting polymers are made from monomers and dopant, which polymerise during oxidation process.[62] The oxidation can be induced by chemical oxidising agents or by applying an anodic current through monomers in the electrochemical cell. The conducting polymer serves as suitable substrate for immobilisation of bio-sensing enzyme such as glucose oxidase and lactate oxidase through adsorption, covalent binding, and crosslinking.[63] In non-enzymatic bio-sensor, the intrinsically conducting polymers increases the charge transfer reaction enhancing the electro-oxidation of glucose by metal oxides.[64] The charge transfer in intrinsically conducting polymers uses the π orbital interactions of conjugated polymer while σ bonds preserve the chain structure. The electrons jump from π orbital in valence band known as highest occupied molecular orbital (HOMO) to the π* orbital in conducting band namely lowest unoccupied molecular orbital (LUMO). The dopant reduces the energy gap between highest occupied molecular orbital (HOMO) and lowest unoccupied molecular orbital (LUMO). The utilisation of polymers for chemical and physical sensing in wearable devices is indicated in Fig. 5.

### Type of Wearable Sensors Based on Polymer and Their Performance as Sensors

Wearable sensors can be categorised according to their functions, physical signals, and chemical constituents. Here, we have classified wearable sensors according to their design or morphology because comfort is highly correlated with the design and materials.[65,66] Being inspired to conform with human skin, the development of wearable sensors is changing from patch-like to thinner, skin-like sensors, which look like a tattoo. Such sensors can be soft, stretchable, and responsive sensors. The challenge is to synthesise a breathable, durable, ultra-sensitive sensor with many functionalities in such very thin sensor. Alternatively, sensors can be added into our daily wearables or accessories such as clothes, contact lenses, rings, and gloves. The merits and disadvantages of each type of sensor are shown in Table II.

In this review, tattoo-type, patch-type, textile-based and other sensor types based on daily wearable devices are discussed. The fabrication method, materials, and sensing performance are outlined.

*Tattoo-type sensors.*—Tattoo-like sensors have sub-micron thickness and usually serve for continuous monitoring over a certain period. Learning from the conventional thick patch-type sensors that can be detached due to bending and stretching of the skin (up to 30%),[81] the tattoo-like sensor should be stretchable and conform to the human skin, acting as an artificial, "second" skin. When the thickness of the tattoo is smaller than the natural wrinkles of the skin (15–100 $\mu$m) the user may not be able to discern the artificial skin.[82] The sensor material should be breathable, allowing perspiration from the skin surface to cool the body temperature.

Like the aesthetic tattoo, tattoo-like sensors can be applied using an invasive method[83] or a non-invasive method via an adherent tattoo sticker.[81] The invasive method utilises the colour change or an optical variable ink achieved by the use of chromogenic dyes, fluorescence, diffraction grating, and plasmons injected within the dermis.[83,84] This approach has been evaluated in ex-vivo porcine skin tissue for sensing pH, glucose (in mixtures with glucose oxidase and hydrogen peroxide) and albumin. However, the reversibility of glucose and albumin detection needs to be improved in order to be used as permanent tattoo-like sensor. Non-enzymatic reversible glucose sensor can be realised by phenylboronic acid with $\beta$-cyclodextrin,[85] which able to reversibly bind the cis-diol of glucose molecules.[86]

Recently, most research has been carried out non-invasive tattoo-like sensors. Earlier, costly microfabrication processes were used for patterning the surface of the thin electronic tattoo, which usually involves high-cost equipment and cleanroom laboratories.[82] Nowadays, the cost-effective methods such as screen printing,[81,87] "cut and paste" method[88] and direct writing[11,89] and have been developed for fabricating the tattoo in Fig. 6. Screen printing uses inking through a patterned stencil to create sensor electrodes, cut and paste utilises cutter plotter to draw the electrode followed by removing extraneous part, while direct writing designs the electrode using inkjet or direct ink printing. The electrode sticks to adhesive layer which also functions as cover to protect the sensor from the environment. The signal from electrode is processed in signal processing unit which can be directly interpreted qualitatively or digitised in quantitative analysis. Colour changing or optical variable ink such as chromogenic dyes, fluorescence, diffraction grating, and plasmons can be used for qualitative analysis. For quantitative measurement, the signal is translated by transducer, amplified, and filtered to reduce noise. The data can be digitised and wirelessly transferred by transceiver. A flexible printed circuit board (PCB) made of screen-printed silver ink on polymer, such as polyimide, polyethylene terephthalate (PET), and polyethylene naphthalate (PEN), is usually used for signal processing circuit. The integration of electrode, signal processing unit, transceiver, and power system in a thin tattoo-like sensor becomes a challenge for researcher in the recent years.

Primarily, the tattoo-like sensor can be categorised in three main functions that are chemical/electrochemical,[81] thermal,[90] and strain sensor.[91,92] Contrary to the invasive tattoo which detect skin interstitial fluid, tattoo stickers for electrochemical sensor can obtained information from sweat, tears, and saliva. Sweat is biofluid which can be accessed on epidermis containing a wealth of biomarker such as electrolytes (e.g., pH, sodium, and potassium ion) and metabolites (e.g., glucose, lactate/lactic acid). The sweat pH contains information such as hydration level as the $Na^+$ is depleted during dehydration as well as an indicator for body odour and skin diseases.[73] Based on a Nernstian response, the pH sensitivity is limited to about 59 mV pH$^{-1}$ at 25 °C, based on the maximum change in anodic and cathodic potentials.[93] Electrochemical sensors based on conducting polymer usually follow the Nernst equation with the sensitivity value about 40–50 mV · pH$^{-1}$ at 25 °C with Ag/AgCl reference electrode stabilised by a KCl-saturated insulator.[81] Ion recognition sites are synthesised during oxidation of conducting polymers exhibiting a pH response. Among the conducting polymers used, polypyrrole (PPy) and polyaniline (Pani) have the highest pH sensitivity, which are 43.2 mV pH$^{-1}$ and 50.1 mV pH$^{-1}$ for PPy[68] and Pani,[67] respectively.

The sensitivity of field-effect transistors (FETs)[93] and charge-coupled devices (CCDs)[94] may exceed the Nernstian limit (59 mV pH$^{-1}$). This could be achieved by accumulation of counter ions above critical potential causing crowding or steric effect of ions near the surface rather than following a classical Boltzmann model of ion distribution.[95] Such a crowding effect could be obtained using extended or dual gate ion sensitive field effect transistor due to capacitive coupling of top and bottom gate which the sensitivity increases as the ratio of top and bottom gate capacitance increases.[93,96–98] The same strategy can be used in a charge coupled device used in a wearable pH sensor.[94] Electrons are transferred from input gate voltage to pH sensing well when a positive bias voltage is switch on. When the positive bias applied to transfer gate, the input gate voltage return to normal stage while the electrons from pH sensing well migrate to capacitor and accumulate.[99] After 100 cycles of accumulation, the sensitivity reaches to 240 mV pH$^{-1}$ in the range of pH 2.8 to 7.1 using silicon oxide as pH sensing layer, PET film as a substrate and polyimide as a cover.[94]

The glucose and lactate detections commonly exploit enzyme such as glucose oxidase and lactate oxidase. The glucose and lactate react with glucose oxidase and lactate oxidase respectively producing hydrogen peroxide which can be detected by electrochemical sensor.[100,101] Alcohol in sweat also can be detected using alcohol oxidase enzyme which shows high correlation of current response with blood alcohol concentration (BAC) at 102 $\mu$A BAC%$^{-1}$.[20] The measurement was faster (less than 60 s) than alcohol detection by transdermal ethanol vapour which has 0.5–2 h delay.[18] In order to provide long term stability, the enzyme should be kept in restricted temperature and humidity ranges, which is inconvenient for highly active people such as athletes. This drawback inspires the development of non-enzymatic biosensor which is based on the reversible reaction of glucose binding. The boronic acid groups are able to reversibly bind 1,2- and 1,3-diol functionalities which are an element of glucose and lactate.[102] Boronate-functional group can be imparted in polymer such as polyaniline[103] and poly(3-aminophenyl)boronic acid[102] as well as in metal oxides such as titanium dioxide.[104] Several redox active metal oxides and hydroxides such as Fe, Cu, and Ni based also has the ability to reversibly bind glucose[105] and lactate[106] through electrochemically and chemically induced reversible redox reactions.

In a hot environment, heat stroke may occur when the body cooling system is not fully functional. Staying hydrated is one of the keys to prevent heat stroke. Wearable temperature sensor can be used as early warning of heat stroke and heat related illness which reminds us to stay hydrated. Having a thin morphology with high conductivity, graphene becomes a popular material for tattoo-like temperature sensor. Lu et al. grew graphene on copper foil and attached to Kapton$^{TM}$ polyimide to be transferred via copper etching.[69] It was stuck on thermal realising tape followed by cutting, heating, and peeling to remove the excess leaving patterned graphene electrode on Tegaderm$^{TM}$ (transparent film dressing) as support. The sub-micron thick graphene sensor was able to withstand 15% stretching up to 1300 cycles and had the temperature coefficient of resistance of 0.0042 °C$^{-1}$, which is comparable to commercial thermocouple.[69] The durability of graphene sensor can be improved significantly using an ionic crosslinked silk as support resulting in high durability over 10000 cycles of 50% strain.[92] The toughness caused by hydrophobic - hydrophobic and ionic interactions between graphene and silk via calcium ions improving shear strength of layered graphene.[107] The non-covalent interaction also gave self-healing ability with 100% healing under 0.3 s.[92] To support the temperature sensor, another functionalities such as UV detection can be added to skin-like sensor by using azobenzene, which its sheet resistance is highly influenced by the intensity of UV irradiation.[108]

The high sensitivity of the tattoo like strain sensor enables it used for respiration monitoring. Recently, the Japanese art of paper crafting such as "Kirigami" and "Origami" give a huge influence in the design of two-dimensional strain sensor. The term "kirigami" comes from the Japanese words "kiri" cut and "gami" paper, which can be translated as the art of paper cutting. Meanwhile, "ori" in "origami" means folding, i.e., it involves paper folding rather than cutting. These paper crafting techniques have been used in space engineer to transport a solar panel in a small packaging.[109] The ability to transform a rigid planar material into an agile design which can be fold and unfold repeatedly without over-stretching the material is beneficial in the strain sensor. There are three methods (e.g., island-bridge, accordion, and kirigami) of paper crafting techniques which is commonly used in two-dimensional strain sensor in Fig. 7.[110] The kirigami design enables the strain sensor to maintain its sensitivity (>80%) at 60% strain as well as more than 60000 cycles of respiration without degradation in performances.[70]

**Table II.** Advantages and disadvantages of sensor types and typical performance.

| Type | Advantages and Disadvantages | Performance (mechanism, active material, substrate) |
|---|---|---|
| Tattoo | **Advantages:** -Conformable to skin. | **pH:** sensitivity = 50.1 mV pH$^{-1}$ (electrochemical, Pani, adhesive polymer)[67]; sensitivity = 43.2 mV pH$^{-1}$ (electrochemical, PPy).[68] |
| | **Disadvantages:** -It may need re-calibration. | **Lactate:** sensitivity = 644.2 nA mM$^{-1}$ and limit of detection (LoD) = 1 mM (chronoamperometric, lactate oxidase, polyethylene terephthalate (PET))[25] |
| | -It can be rubbed off with water. | **Glucose:** sensitivity = 23 nA $\mu$M$^{-1}$; LoD = 3 $\mu$M (chronoamperometric, glucose oxidase, tattoo paper)[19] |
| | | **Alcohol:** sensitivity = 102 $\mu$A BAC%$^{-1}$ (electrochemical, alcohol oxidase, gel).[20] |
| | | **Temperature:** sensitivity = 0.0042 °C$^{-1}$ (resistivity, graphene, acrylic polymer).[69] |
| | | **Respiration:** >60000 cycles at 60% strain with gauge factor (GF) ≈1.33 (resistive, graphene, polyimide).[70] |
| Patch | **Advantages:** -Reliable and consistent measurement for continuous monitoring. | **Tyophylline:** maximum limiting current ($I_{max}$) = 0.31 $\mu$A and Michaelis-Menten constant ($K_M$) = 13 mM (chronoamperometric, xanthine oxidase, polycarbonate)[71] |
| | | **Lactate:** $I_{max}$ = 0.95 $\mu$A, $K_M$ = 0.7 mM (chronoamperometric, lactate oxidase, polycarbonate)[71] |
| | **Disadvantages:** -Cannot withstand stretching. | **Glucose:** $I_{max}$ = 23 $\mu$A, $K_M$ ≈ 13 mM (chronoamperometric, glucose oxidase, polycarbonate)[71] |
| | | **Levodopa:** sensitivity = 0.038 nA $\mu$M$^{-1}$; LoD = 0.25 $\mu$M (chronoamperometric, tyrosinase mushroom enzyme, Nafion)[72]; sensitivity = 0.082 $\mu$A $\mu$M$^{-1}$; LoD = 0.5 $\mu$M (square wave voltammogram, graphene and mineral oil, Nafion).[72] |
| Textile | **Advantages:** -Can be adapted into daily apparel. | **pH:** sensitivity = 0.1 (colorimetric, pH sensitive dyes, polymer gel).[73] |
| | -Comfortable for highly active user such as athlete. | **Lactate:** range of detection = 0–12.5 mM (colorimetric, lactate dehydrogenase, hydrogel).[74] |
| | | **Influenza protein:** range of detection = 10 ng mL$^{-1}$ – 10 $\mu$g mL$^{-1}$ (electrochemical, functionalised carbon, polyamide).[75] |
| | | **Strain:** >10000 cycles at 100% strain with gauge factor (GF) ≈0.68 (resistive, carbon, PDMS)[76]; |
| | **Disadvantages:** -Lack of intimate contact. | >300000 cycles at 40% strain with GF ≈0.65 (resistive, carbon nanotubes, polyurethane)[77]; |
| | -Washing degradation. | 2000 cycles at 14% strain with maximum GF ≈6.02 (capacitive, MXenes, cotton).[78] |
| Contact lenses | **Advantages:** -Minimal disruption | **Glucose:** minimum detectable concentration 12.57 $\mu$M (transistor, graphene, EFiRON™)[79]; |
| | | minimum detectable concentration 0.4 $\mu$M (transistor, graphene-AgNW, Ecoflex™)[22] |
| | **Disadvantages:** -Red eye and irritation may occur | **Intraocular pressure:** sensitivity = 109 $\mu$V mmHg$^{-1}$ (resistive, Pt/Ti strain gauge, silicone elastomer)[80]; |
| | | sensitivity = 2.64 MHz mmHg$^{-1}$ (capacitive, silicone elastomer sandwich with graphene/Ag, silicone elastomer)[22] |

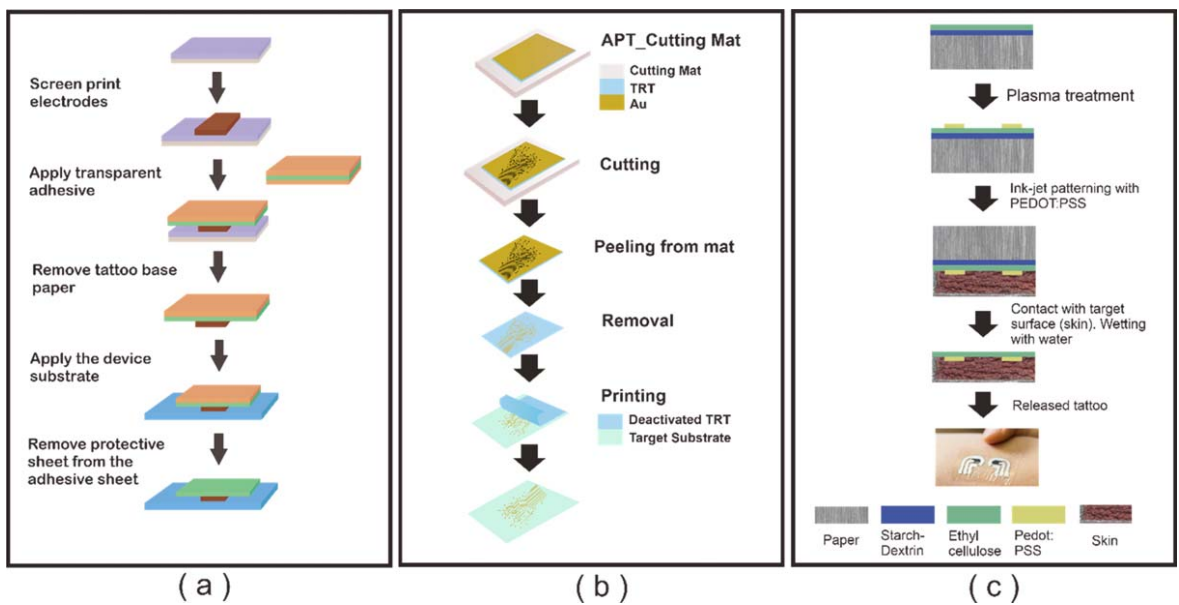

**Figure 6.** Several methods to fabricate tattoo-like sensor. (a). screen printing.[43] (b). "cut and paste".[88] (c). direct writing.[11]

*Patch-type sensors.*—While some researchers focus on fabricating thinner patch-type sensor, the others pursue better interfacial contact between electrode and human body.[111] Polymer is commonly used for electrolyte gel in wet electrode. However, the gel may cause skin irritation and discomfort. Dry contact or capacitive electrode become a potential candidate to replace it. Dry capacitive electrode is a non-contact electrode which has a layer of insulator between electrode and skin. Although it completely eliminates the skin irritation, the signal to noise ratio is low due to high impedance between skin and sensor as well as high sensitivity to mechanical/movement noise.[112] In order to provide better contact without the gel, researchers fabricate microneedle pattern thus the electrode may reach the epidermis in Fig. 8. Stratum corneum act as an information barrier creating noise in the measurement. The outer surface of skin can be contaminated by environment or dead skin cells, the skin motion may loosen the electrode contact. Also, the high impedance at the interface of electrode and skin hampers the signal source. The needles need to be strong enough to puncture the stratum corneum (10–20 $\mu$m in thickness) and penetrate to the epidermis (50–100 $\mu$m) to collect biofluid. The impedance of skin can be interpreted using circuit model of parallel resistors ($R$) and capacitors ($C$) where

$E_{el}$ is the half-cell potential of electrode
$C_{el}$ and $R_{el}$ represent electrode-electrolyte impedance
$R_{gel}$ is the resistance of electrolyte gel
$E_{sc}$ is the half-cell potential of electrolyte-skin interface
$C_{sc}$ and $R_{sc}$ represent the impedance of skin
$R_{sg}$ is the resistance of subcutaneous tissue and dermis.

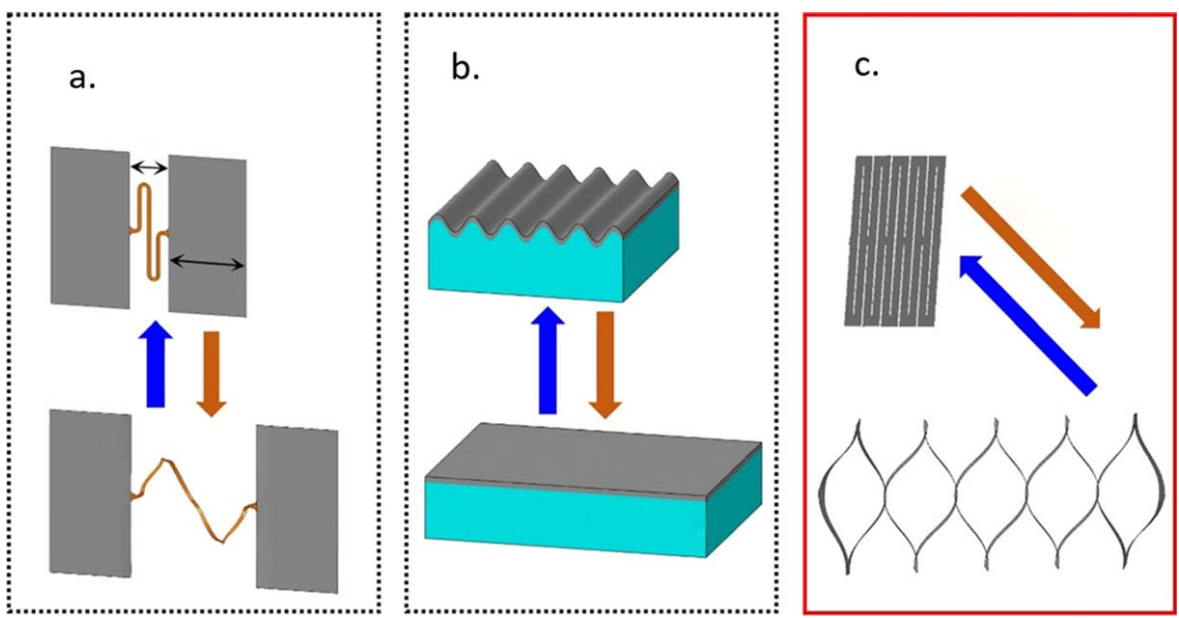

**Figure 7.** Paper crafting techniques for a two-dimensional strain sensor. (a). Island-bridge, (b). accordion, (c). kirigami.[110]

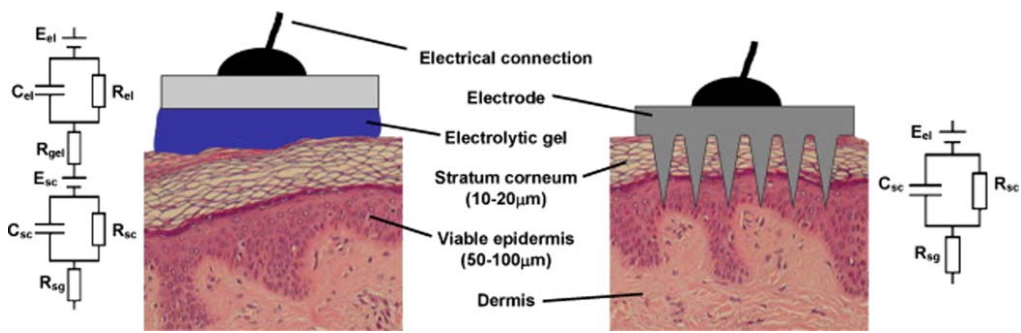

**Figure 8.** Electrical model and sketch of wet (left) and microneedles dry electrode (right) contact on skin surfaces. The electrical model shows additional resistance of gel in wet electrode while direct contact in microneedles dry electrode reducing resistance at the interface.[113]

In wet electrode, the electric current is hampered at the interface of electrode to gel and gel to skin which is weaken the signal to noise ratio, while direct connection of dry electrode and skin removes the resistance at the interface. In terms of the material, metal, although it is conductive, it is difficult to stretch. Conductive elastomers are the most promising candidates for this purpose.

To fabricate a microneedle array on a conductive elastomer, photo-lithography,[114] injection moulding,[71] soft lithography (e.g., replica moulding[115] and embossing[116,117]), and 3D printing[118] are commonly used. In photo-lithography, the material surface is covered with photo-resist materials and is irradiated with masked ultraviolet followed by etching of the photoresist material developing desirable pattern on the surface. This process requires sophisticated equipment and materials. Contrary to photo-lithography, the first step in injection moulding is to create patterned negative mould which usually deploys electric discharge milling (EDM) or laser ablation followed by injection moulding of melted polymer and cooling down. In soft-lithography, a micro/nanopatterned elastomer is deployed as replica mould of patterned master surface to minimise the high cost of photolithography.[119] The replica mould could also serve as stamp to create recessed relief pattern on the subject surface. The recent development in 3D printing techniques especially fused deposition modelling (FDM)[120] and stereolithography[121,122] enabling high resolution polymer printing up to micron size. Metal sputtering, conducting polymer or carbon deposition can be used afterwards to increase the electrical conductivity of polymer microneedles. In case of using soft elastomer such as polydimethylsiloxane (PDMS), the geometry of PDMS pillars is important to prevent pairing or lateral collapse in Fig. 9.[123] As a rule of thumb, the length ($H$) to width ($L$) ratio of pillar should be more than 0.5 and lower than 5 ($0.5 < H/L < 5$) also the ratio of length ($H$) to distance between pillar ($D$) should be more than 0.05 ($H/D > 0.05$).

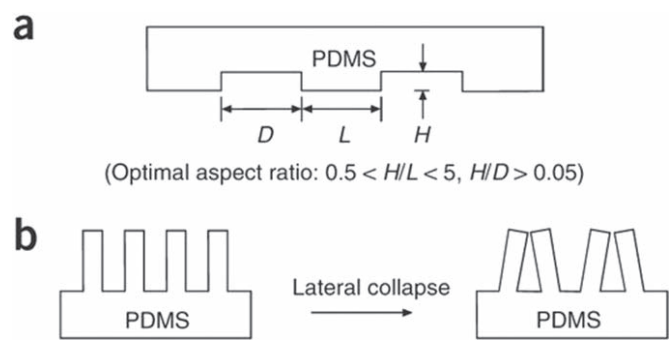

**Figure 9.** A schematic of a PDMS microneedle array geometry. (a). the parameters in PDMS microneedle array geometry in which $H$, $L$, and $D$ are the height, diameter, and distance between needles, respectively. (b). the optimal aspect ratio is shown in order to prevent several problems arise due to the softness of PDMS such as lateral collapse or pairing ($H/L > 5$).[123]

Instead of sensor applications, microneedle arrays are also utilised for transdermal delivery of drug or vaccine which is a minimally invasive drug or vaccine injection.[124–126] The microneedles may be encapsulated with drug or to be used for drug infusion into the skin. This technique can be ultimately combined with microneedles sensor creating a sensor that can administer drug at the same time or to monitor the drug delivery process. Wang's group at University of California has used microneedles sensor for continuous monitoring of levodopa (L-dopa) which is a medication for Parkinson's disease (PD).[72] The symptom of Parkinson's diseases occurs when the dopamine level is decrease. Levodopa replenishes dopamine while it crosses the blood brain barrier.[127] Hence, continuous monitoring of levodopa is crucial to prevent dopamine wearing off in the patient with PD. The electrochemical microneedle sensor utilised enzymatic and non-enzymatic electrodes.[72] The non-enzymatic needle electrode was made of graphite and mineral oil and the enzymatic electrode used tyrosinase mushroom enzyme which immobilised on the surface of graphite and mineral oil as an active material to detect transition from levodopa to dopaquinone during electrochemical measurement. The electrode was covered by Nafion$^{TM}$ (e.g., ionically conductive polymer) to cover the electrode from contamination. The reference needle electrode was modified with Ag. The anodic detection of L-dopa was achieved using non-enzymatic needle through square-wave voltammograms (SWV) while the chronoamperometric detection was performed using enzymatic needle. The sensitivity could reach up to 0.038 nA $\mu M^{-1}$ and 0.082 $\mu A$ $\mu M^{-1}$ for the chronoamperometric and SWV, respectively. The limit of detection (LoD) for chronoamperometric and SWV was 0.25 $\mu M$ and 0.5 $\mu M$, respectively.

The microneedles sensor has shown to effective in continuous monitoring of glucose, lactate, and tyophylline as well.[71] Tyophyilline is a known drug for respiratory diseases such as asthma and chronic obstructive pulmonary disease.[128] Cass et al. was utilised aluminium mould patterned with electrical discharge milling (EDM) to be used for injection moulding of polycarbonate microneedles.[71] To improve the electrical conductivity, the chromium (15 nm) and platinum (50 nm) were sputtered on polycarbonate microneedles as working electrode. For reference electrode, the microneedles were sputtered with Ag (150 nm) and treated with $FeCl_3$. The working electrode was decorated with specific enzyme such as glucose oxidase, lactate oxidase, and xanthine oxidase for chronoamperometric measurement of glucose, lactate, and tyophylline, respectively. In which the maximum limiting current ($I_{max}$) and Michaelis-Menten constant ($K_M$) were ≈23 $\mu A$, 0.95 $\mu A$, 0.31 $\mu A$ and ≈13 mM, 0.7 mM, 13 mM, respectively for glucose, lactate, and tyophylline sensor.

*Textile-type sensors.*—Based on the working principles, textile-based sensor can be categorised into electrochemical, transistor-based, and passive stimulus-response sensor in Fig. 10.[129] Three electrodes configuration such as reference, working, and counter electrode are sewn on the textile and are used for electrochemical sensor with sweat as electrolyte. Carbon or Ag/AgCl conductive ink is usually deposited on textile as electrodes by screen printing[75] or dip coating.[130] Carbon based electrode such as graphene can be utilised as substrate for biomarkers such as 1-Pyrenebutyric acid-N-hydrosuccinimide ester (PANHS), which is an active material for binding influenza A-specific antibody.[75] The bound and unbound reaction occurs during electrochemical impedance measurement. The resulting sensor was able to detect the influenza protein as low as $10 \, \mathrm{ng \, mL^{-1}}$ up to $10 \, \mathrm{\mu g \cdot mL^{-1}}$. In a transistor sensor, the electrons are passed from source to drain through gate and channel in which the channel conductivity is controlled by gate. The voltage can be applied to gate electrode inducing reaction in the channel with molecules in sweat. The conducting polymers such as poly(3,4-ethylenedioxythiophene):polystyrene sulfonate (PEDOT:PSS), polypyrrole (PPy), and polyaniline (Pani) are usually used as channel and modified by enzyme (e.g., glucose oxidase and dehydrogenase)[131] or redox active analytes (e.g., adrenaline, dopamine and ascorbic acid),[132] in which the redox reactions occurs during applied voltage.

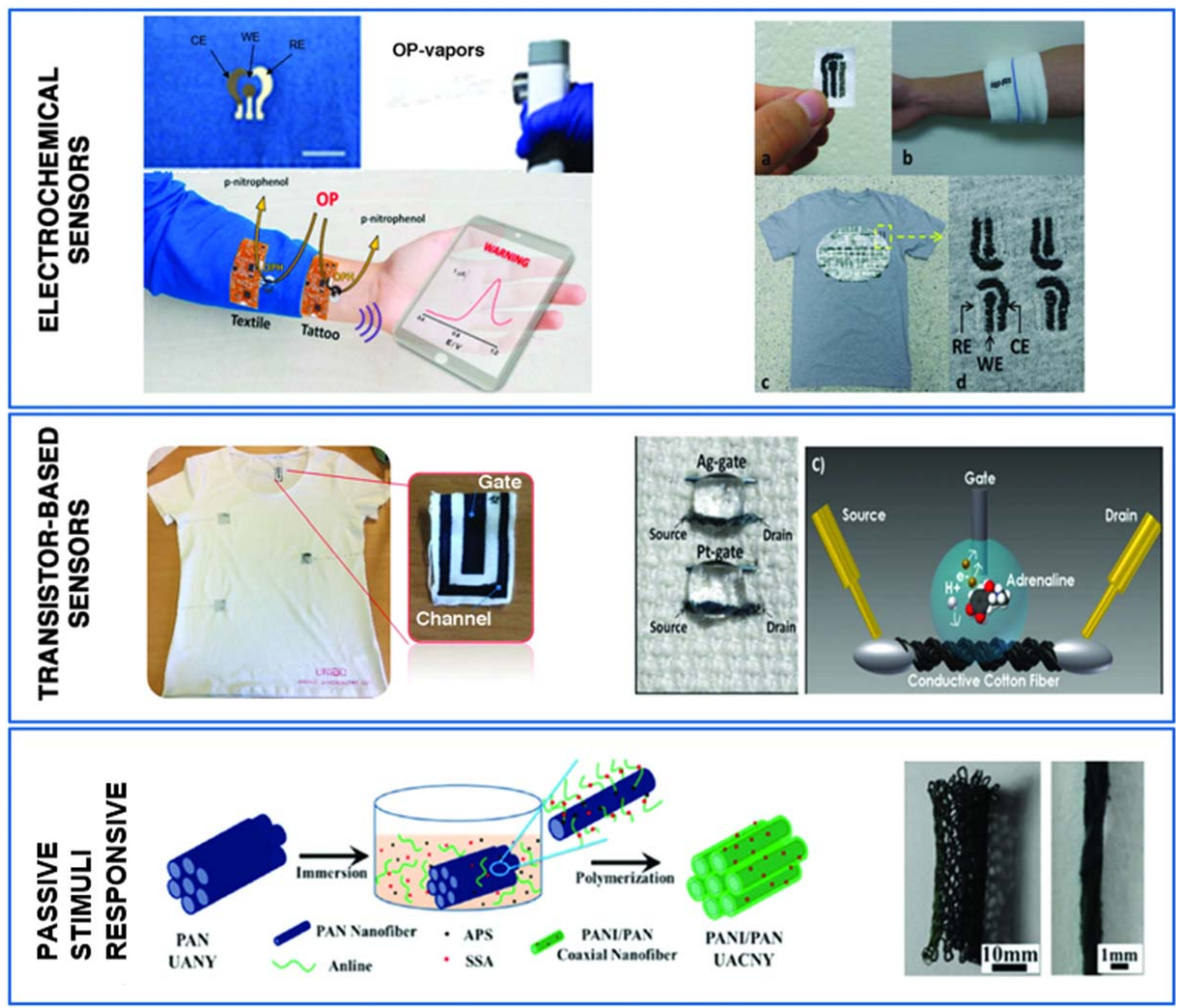

**Figure 10.** Three types of textile-based sensor. Electrochemical sensors: three electrode configurations for detecting organophosphate vapour, glucose and lactate. Transistor based sensor: oxidation of adrenaline at conducting fibre channel is detected. On a passive stimulus-response sensor, fibre can be coated by conducting polymer which can modify resistivity in the presence of chemical such as ammonia.[129]

In passive stimulus-response polymer sensors, the sensing method is deployed using the observation of direct changes in materials such as resistance, optical, capacitance, which is commonly used to detect strain and chemical sensing.[74,133,134] As a smart textile is often used for highly active user such as athlete, minimal modification is sought to retain its comfort. Hence, power-free sensor such as stimulus-response sensor is preferable. A colorimetric dye is one of the most common methods for passive detection of chemical in sweat.[74] The pH of human sweat is typically between 4.5 and 7, so pH sensitive dyes (e.g., bromocresol green, methyl orange, bromothymol blue, and bromocresol purple) should exhibit a colour change in this pH range.[67] To contain the dyes within the textiles, hydrogel can be used with surfactant to prevent swelling of hydrogel due to humidity. Alternatively, ionic gel is more stable to humidity but one must note the anions or cations contain in the gel may interfere the pH indicator.[73] For lactate detection, the purple lactate colour intensity increases as the lactate concentration increases due to enzyme reaction of lactate dehydrogenase, where changes can be observed up to 12.5 mM as the threshold concentration.[74] As an alternative to a colorimetric dye, a plastisol based microfluidic channel can be used to detect chemical such as sodium ions.[135] The plastisol resistivity is sensitive to the concentration of sodium ions. The microfluidic channel offers focused detection of the chemical solution considering the fabric may absorb the solution as well. Instead of resistivity-based sensor, passive detection can be pursued by capacitive sensor.[13] The humidity can be measured using polyimide which has imide group hence it is able to create hydrogen bonding with water. The capacitance change when the hydrogen ions in water fill the microvoid within the polyimide.

The stimulus-response strain sensor works based on the measurable changes in resistivity, capacitance, piezoelectric, triboelectric, optical during stretching or bending.[134] The resistive strain sensor is the most common type in textile-based sensor which utilises the fabric as resistor and measures the resistivity changes. The electrical conductivity of fibres can be manipulated through coating, spinning, and knitting of conductive materials such as conducting polymer, carbon-based structures (e.g., graphene, carbon nanotubes, carbon black nanoparticles, carbon fibres), and silver. The conducting polymer can be deposited on non-conducting fabric by chemical polymerisation.[136] Typically, the process begins with dipping the fabric into monomer solution containing dopant followed by soaking it into the oxidant at controlled temperature. The opposite can be achieved using vapor phase polymerisation where the fabric is soaked into oxidant and dopant followed by exposing the fabric to the monomer vapour.[137] Alternatively, a one-pot method can be deployed using repetitive dipping of fabric into a dispersed solution of conducting polymer.[138] However, cracking can occur in conducting polymer coated fabrics due to the brittle nature of some conducting polymers.[139] Elastomers, such as PDMS can be used either as an interface material between fabric and conducting material or as a mixture with the conducting composite. The conducting filler in the composite may create a percolative network or conducting path during stretching which increases the conductivity. Carbon/PDMS core/sheath composite strain sensors can retain their performance over more than 10000 cycles at 100% strain with a gauge factor ≈0.68.[76] Carbon nanotube coated cotton/polyurethane core/sheath composites strain sensors have also shown excellent durability with a gauge factor ≈0.65 at 40% strain over more than 300000 cycles.[77] However, the comfort of the textile can be compromised due to the hydrophobic nature of most elastomers. Instead of carbon and conducting polymer, the recent development of MXenes (two-dimensional transition metal carbides) has created high sensitivity capacitance strain sensors with a maximum strain gauge ≈6.02 and it is able to withstand 2000 cycles at ≈14% strain although the maximum strain is only 20%.[78] Such a sensor was made by two step dip coating of cotton in small diameter MXene solution followed by a large diameter MXene solution. For further work, the maximum strain cycles of MXene-coated textiles could be improved by using an elastomeric substrate.

*Contact lens sensors.*—While tattoo, patch, and textile collect data from sweat or saliva, contact lenses obtain information from tears in Fig. 11.[79] The glucose level in tears is averaged 3.6 mg · 100 ml$^{-1}$ (0.2 mM) and 16.6 mg · 100 ml$^{-1}$ (0.92 mM) for normal and diabetic patients respectively[140] which can be a reliable biomarker for diabetes. The challenge is to integrate the power system, rectifier, transmitter, and sensor electrodes within the lenses while maintaining biocompatibilty with delicate human eye tissue.

Another issue is the glucose level from tears has lag time around 10–20 min compared to blood glucose level.[21] Since the aim of glucose sensing contact lenses is continuous glucose monitoring, the lag time problem could be minimised. Optically transparent and biocompatible polymers such as soft silicone elastomer, Ecoflex$^{TM}$ [22] and fluorinated polymer, EFiRON® [79] are often used as substrate and are deposited by silver nanowires (AgNW) followed by patterning using photolithography to fabricate electrodes, rectifier, and wireless circuit. The field effect transistor-based glucose sensor can be utilised in contact lenses type sensor using graphene as active material immobilising glucose oxide enzyme with a mixed of Cu/Au as source and drain electrodes. The circuit can be connected to small light emitting diode (LED) as indicator although some people argue that the LED may contain toxic arsenic and lead.[141] Using only a few layers of graphene, a transparency of >91% can be retained.[22,79] Chen et al. utilised phenylboronic acid (PBA) based hydroxyethyl methacrylate (HEMA) contact lens to fabricate a non-enzymatic glucose sensor.[142] The sensing mechanism was done by measuring the changes in width and thickness due to swelling during glucose interaction with boronic acid. Although cytotoxicity studies showed no significant difference in cell viability, the comfort of the lenses is questionable due to swelling. Besides transistor-based and swelling based contact lenses sensors, an electrochemical tears sensor has been developed, particularly the NovioSense Glucose Sensor, which has been through phase II clinical trials with 6 patients with type 1 Diabetes Mellitus.[143] The tears sensor contained a flexible coil (diameter of 60 μm) containing working, counter, and reference electrodes as well as transmitter. The outer layer of coil was made of hydrophilic polysaccharide immobilised with glucose oxide enzyme. The coil was then put on the lower eye lid for chronoamperometric measurement. The performance of NovioSense Glucose Sensor was similar to Abbott FreeStyle Libre, blood glucose sensor.

The pressure or strain sensor in contact lenses can be utilised to detect the intraocular pressure (IOP) of bovine eyeballs which is useful for glaucoma patients (see Fig. 12[22]). Tonometry is commonly used to measure intraocular pressure of glaucoma patients although it has many weaknesses such as single point and one-time (rather than continuous) measurement not to mention the discomfort when a tonometer probe applies pressure to the eyes.[144] As an alternative, a wireless non-invasive pressure sensitive contact lenses sensor can be deployed either using a resistive-based[80] or capacitive-based[22,145] sensor. In resistive-based sensors, two thin Pt/Ti strain gauges have been microfabricated and sandwiched between two polyimide layers. One strain gauge was used to measure changes in the corneal curvature due to IOP and the other placed radially to minimise strain while acted as baseline measuring strain caused by thermal expansion. The gold antenna was then electrodeposited on the polyimide and connected to gauge by conductive epoxy. The gauge, antenna, and wireless microprocessor were embedded on the silicone contact lenses by cast moulding technique. The resulting sensor was tested on pig eyes and showed high linearity and a sensitivity of 109 $\mu$V · mmHg$^{-1}$ between 20 and 30 mmHg, which is the pressure range of glaucoma patient.[80] For capacitive-based sensors, a silicone elastomer, Ecoflex$^{TM}$, was used as dielectric and sandwiched in between a spiral coil of graphene and silver nanowire electrodes.[22] A bovine eyeball has been used for in-vitro measurement and the resonance frequency was calculated as a function of inductance and capacitance with the sensitivity of 2.64 MHz · mmHg$^{-1}$ between 5 to 50 mmHg.[22] These pressure sensitive contact lenses sensors could be combined with drug release of anti-glaucoma drugs, such as timolol, to evaluate the effectiveness of glaucoma treatment. However, calibration with a commercial tonometer may be needed due to the variation of corneal curvature between individuals.

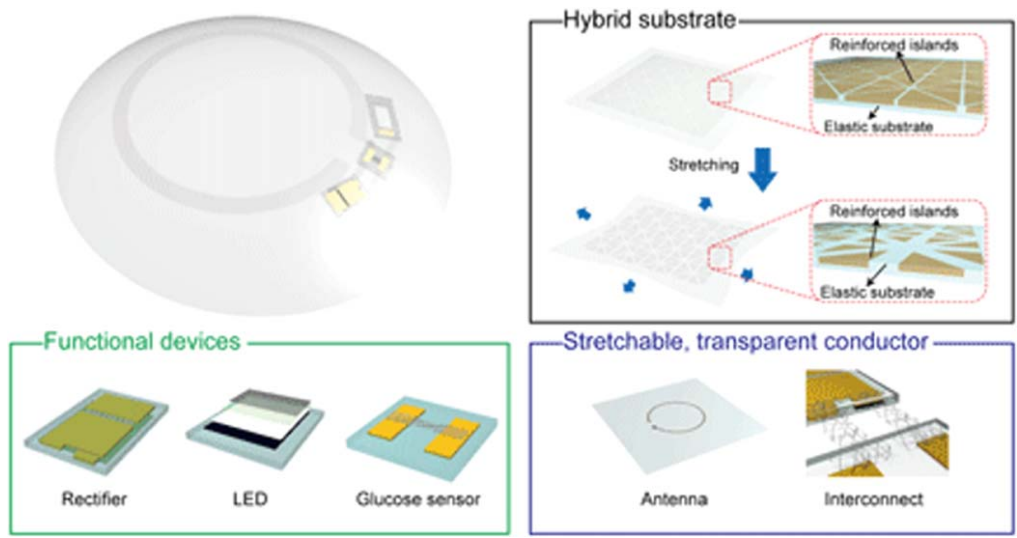

**Figure 11.** Contact lens devices containing rectifier, LED, and glucose sensor using graphene and silver nanowires as conducting materials. The glucose oxidase enzyme can be immobilise on graphene for glucose detection. While the current transfers to the antenna, it transmits a signal to the LED, shutting it off when it reaches a certain glucose level.[79]

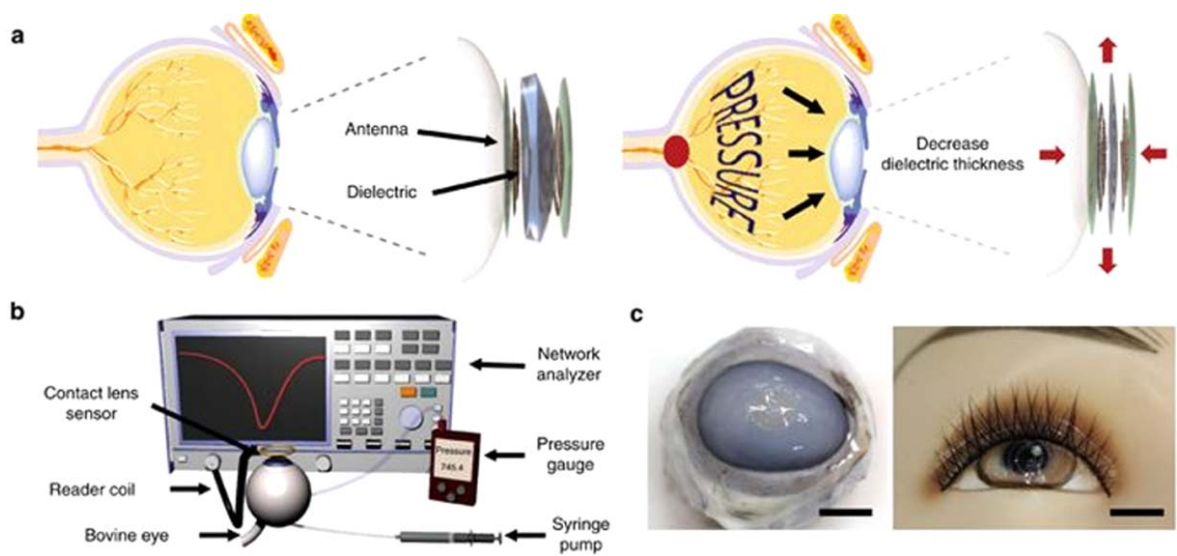

**Figure 12.** A schematic showing intraocular pressure sensing using relative changes in dielectric. (a). The intraocular pressure changes the thickness of the dielectric altering the dielectric value. (b). The set-up of in-vitro testing of the intraocular sensor. (c). The photographs of bovine eyeball (left) and mannequin eye (right) wearing the contact lenses sensor.[22]

### Conclusions

Having the ability to stretch without breaking and chemically resistant, the polymers are mostly used as flexible substrate that may conform to the human body as wearable sensor. The combination of polymer and conducting filler composite is utilised as stretchable dry electrode since the conventional wet electrode may cause problems such as skin irritation and decreasing performances as it dries. The development of intrinsically conducting polymers led its utilisation as dry electrode, strain sensitive sensor, a vessel for enzyme immobilisation, and as catalyst for redox reactions. Such sensitivity of epidermal and tattoo strain sensor can be extended into tactile artificial robotic skin.[146] Electroactive hydrogel may also enable dual functions of device such as actuators and sensors. Ionic gel acts as conductive substrate that is able to shield and contain the chemical reactions from the environment such as humidity. Knowing the uniqueness of polymer such as liquid crystal polymer, stimulus-response polymer gels, and piezoelectric polymers, there is considerable scope to expand the use of these polymers in wearable sensors.

The polymer-based strain sensors mostly rely on resistivity changes of conducting polymer while piezoelectric polymers are receiving less attention, although piezoelectric polymer may produce self-powered sensor.[147] The functional groups of liquid crystal polymer and polymer gel can be tailored for versatile active materials in passive stimulus-response sensors. Using functional stimuli responsive polymer gels, the sensor can be combined with other applications such as imaging[56] and drug delivery.[54] Contrast agent functionalised hydrogel has been widely used for in-vivo imaging, which may inspire researcher to expand its utilisation in non-invasive wearable sensor. Combination of emerging technologies such as organic printed electronic with polymer-based wearable sensor could be an important strategic research direction. The flexible substrate has a very significant role in organic printed electronics as one of the main purpose of printed electronics is to compete with the rigid silicon or circuit based citcuitry.[148] A conducting polymer is necessary as a conductive ink, while durable polymer such as polyimide is a suitable candidate for flexible substrates. Polymers have high processing ease and versatility which can be used in electrospinning[149,150] and 3D printing.[151,152] A challenge is to improve the charge transfer efficiency of active material and analytes hence enhancing the sensitivity or decreasing

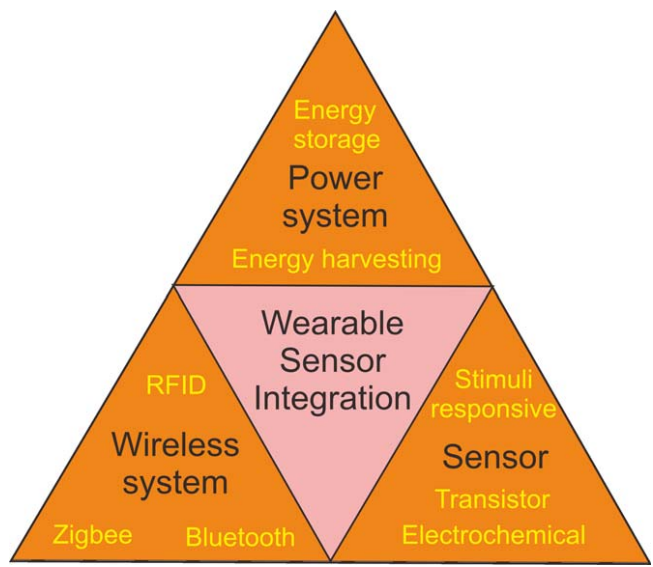

**Figure 13.** The integration of components in a wearable sensor: power system, wireless system and sensor.

delay time of non-invasive sensor, improving device lifetime, and reducing energy consumption.[153]

In recent years, research on wearable devices has moved towards the integration of sensor, wireless, and power systems[154] as depicted in Fig. 13. Since the wearable sensors should be lightweight and simple, the low energy wireless system such as Zigbee, radio-frequency identification (RFID), and low energy Bluetooth are preferable. The power system of wearable devices can be pursued using wearable batteries,[15] supercapacitor,[155] and bio-battery/fuel cell[156] for energy storage while wearable photovoltaic[155] and kinetic energy harvester (e.g., piezoelectric, electrostatic, and electromagnetic)[157] for energy harvesting. The wearable batteries have been commercially available based on zinc air batteries for hearing aid. It has disadvantage such as non-rechargeable and diminishing performance due to evaporation of moist electrolyte. Many researchers focused on all solid-state lithium ion batteries which can be stable for very long time and rechargeable. A solid polymer gel electrolyte can replace a conventional liquid electrolyte and act as a binder for flexible electrodes.[158,159] In wearable supercapacitor, polymer acts not only as solid electrolyte but also electrodes and separator.[160] The conducting polymer possess high theoretical specific capacitance which is suitable for pseudocapacitor.[161,162]

Some studies pursue self-powered wearable energy storage by deploying redox reaction of carbohydrate or glucose as fuel.[156] Redox polymer binds and coats the enzyme then transferring the electron from the enzyme active sites to the electrode.[163] In photovoltaics, a flexible substrate such as polyimide or polyester is important to allow the silicon cell to have some flexibility.[164,165] Ideally, the thin film solar cell electrode material, such as cadmium telluride (CdTe) and copper gallium indium diselenide (CIGS), which can show >20% cell efficiency on a laboratory scale, should be used for wearable photovoltaic instead of silicon.[166] However, the high efficiency thin film solar cells do not meet the cost-efficiency requirement for marketable product. Alternative material such as copper zinc tin sulphide (CZTS) is sought in order to reduce the cost of thin film solar cells. For kinetic energy harvester, piezoelectric polymer such as polyvinylidene difluoride (PVDF) is preferable than ceramic piezoelectric (e.g., lead zirconate titanate (LZT) and barium titanate) due to its cost efficiency, faster synthesis process, and more flexible.[167] In summary, the polymers have indeed made a huge contribution in wearable sensors and its power systems. Further applications of polymers in integrated wearable sensor are anticipated considering the benefits of such devices. However, considerable challenges face designers. On the one hand, devices need to utilise polymers having tailored electroactive and other sensor properties while maintaining sufficient resistance to biofluids (or the capability of using them) to remain active and durable with a high lifetime. On the other hand, the large scale persistence of polymers in our environment necessitates increasing attention to natural, biodegradeable and polymers together with those able to be recycled.